\documentstyle[12pt]{article}
\begin{document}
\begin{center}
\Large\bf The effect of $B\pi$ continuum in the QCD sum rules for the $(0^+, 1^+)$ 
heavy meson doublet in HQET
\end{center}
\vspace{0.5cm}
\begin{center}
{ Shi-Lin Zhu and Yuan-Ben Dai}\\\vspace{3mm}
{Institute of Theoretical Physics \\
Academia Sinica, P.O.Box 2735\\
Beijing 100080, China\\
FAX: 086-10-62562587\\
TEL: 086-10-62541816\\
E-MAIL: zhusl@itp.ac.cn}
\end{center}
\vspace{1.0cm}
\begin{abstract}
We study the effect of $B\pi$ continuum 
in the QCD sum rule analysis of the heavy meson doublet $(0^+, 1^+)$
in the leading order of heavy quark effective theory. New sum rules are derived
for the leading order binding energy ${\bar \Lambda_{+,{1\over 2}} }$ and pionic
coupling constant $g'$.
\end{abstract}


\pagenumbering{arabic}

\section{ Introduction}
\label{sec1} 

The QCD sum rules \cite{svz} for the masses of 
the excited heavy meson doublet 
$(B'_0, B'_1)$ of spin parity $(0^+, 1^+)_{(+, {1\over 2})}$
have been studied in \cite{nucl-shuryak,col-95,dyb-1,zhu-prd}, 
where the indices $(+, {1\over 2})$ denote the parity and spin of 
the light component $j_l$. 
Recently the ${\cal O} (\alpha_s)$ correction to the $m(0^+)$ 
sum rule has been calculated in \cite{ht,col-98}. 
In \cite{dyb-1,zhu-prd} the masses of the 
$(0^+, 1^+)_{(+, {1\over 2})}$, together with those of 
the doublet $(1^+, 2^+)_{(+, {3\over 2})}$ were calculated 
up to the order of ${\cal O}(1/m_Q)$ in the framework of 
heavy quark effective theory (HQET) \cite{grinstein}.
Let ${\bar \Lambda}_{P,j_l} =m_{P,j_l}-m_b$, where $m_{P,j_l}$ is the 
mass of the the doublet in the leading order of $1/m_Q$ and 
$P, j_l$ are the parity and spin of the light component of the 
heavy mesons in the doublet. The results in \cite{zhu-prd} in the leading 
order of $\alpha_s$ are ${\bar \Lambda}_{+, {1\over 2}}=
(1.15\pm 0.10)$GeV if the usual interpolating current of the lowest 
dimension is used and ${\bar \Lambda}_{+, {3\over 2}}=(0.82\pm 0.10)$GeV. 
Within errors the results in \cite{nucl-shuryak,col-95,dyb-1} are 
consistent with these results. This would imply that $0^+$ state lies
$100--300$ MeV above $2^+$ state. The ${\cal O}(1/m_Q)$ corrections
calculated in \cite{zhu-prd} does not change much this mass difference. 
This result is inconsistent with the new experimental 
data \cite{L3} where $m(2^+)$ is about $100$ MeV larger than 
$m(0^+)$. It is unlikely that ${\cal O}(\alpha_s)$ corrections can
account for this discrepancy.

Recently Blok et al. made the following observation \cite{blok}.
Due to the $S$-wave nature of the $B\pi$ intermediate state and 
the large coupling of the soft Goldstone particle, the contribution
of the $B\pi$ continuum to the spectral density in the correlator
of two $0^+$ currents is unusually large. It rises faster than 
the quark-gluon spectral density in the low energy region and 
exceeds it in magnitude in that region. Thus, it may violate the 
naive quark-hadron duality if we integrate the spectra over 
a region below the lowest pole. It was proposed in \cite{blok} 
that this is the reason for the abnormal large value of the residue 
of the pole \cite{nucl-shuryak} obtained in the standard 
"lowest pole plus parton-like continuum model" for the QCD
sum rules. Therefore, a better approach is to include the 
$B\pi$ continuum in the soft pion region in addition to 
the lowest $B'_0$ pole in the sum rule.

In this note we shall study in detail this effect in the sum rules 
for ${\bar \Lambda}_{+, {1\over 2}}$ and the residue of the $B'_0$
pole. Besides, the decay rates of $B'_0$ and $B'_1$ have been 
studied in \cite{col-95,zhu-epjc} with the light cone QCD sum rules 
(LCQSR) \cite{bely-95}. In view of the duality violation effect
due to the intermediate states of $B\pi$ continuum in the sum rule
we shall also re-investigate the LCQSR for pion coupling of 
the $(0^+, 1^+)$ doublet including the contribution of $B\pi$ states.
Section \ref{sec2} is a short review of previous 
sum rules. New QCD sum rules with intermediate state contribution and the 
numerical analyses are presented in section \ref{sec3}. The last section 
is a short summary.

\section{Previous sum rules}

\label{sec2}

\subsection{Previous mass sum rules}

The interpolating current for the doublets $(0^+,1^+)$ reads
\begin{equation}
\label{curr1}
J^{\dag}_{0,+,{1\over 2}}=\bar h_vq\;,
\end{equation}
\begin{equation}
\label{curr2}
J^{\dag\alpha}_{1,+,{1\over 2}}=\bar h_v\gamma^5\gamma^{\alpha}_tq\;,
\end{equation}
where $h_v (x)$ is the heavy quark field in HQET, $v_\mu$ is 
the heavy hadron velocity, $\gamma_t^\alpha =\gamma^\alpha -{\hat v}v^\alpha$,
the indices $j, +, j_l$ in $J_{j,+,j_l}$ are the total angular momentum, 
the parity and the light component angular momentum respectively.
Note there is a factor of ${1\over \sqrt{2}}$ in the definitions of the 
interpolating currents in our previous work. 

We consider the correlator 
\begin{equation}\label{cor-1}
\Pi (\omega )=i\int d^4 x e^{ikx} \langle 0| T\{J_{0,+,{1\over 2}}(x), 
J^{\dag}_{0,+,{1\over 2}} (0)\} |0\rangle \; 
\end{equation}
and define the overlapping amplitude $f_{+,{1\over 2}}$ as
\begin{equation}\label{overlap}
\langle 0 | J_{0,+,{1\over 2}} (0) | B'_0\rangle =f_{+,{1\over 2}} \; .
\end{equation}

Assuming the standard "pole plus parton-like continuum" model we get
\begin{equation}
\Pi (\omega )={1\over 2} {f^2_{+,{1\over 2}} 
\over {\bar \Lambda}_{+,{1\over 2}}-\omega}
+\mbox{continuum} \;.
\end{equation}
Here $\omega =v\cdot k$, which is a factor ${1\over 2}$ smaller 
than the $\omega$ used in \cite{zhu-prd}.
We derive the sum rule for the $(0^+, 1^+)$ doublet \cite{dyb-1}
\begin{equation}
\label{mass-0}
{1\over 2}f_{+,1/2}^2e^{-{\bar\Lambda_{+,{1\over 2}}/T}}=
\frac{3}{2\pi^2}\int_0^{s_0}\omega^2e^{-\omega/{T}}d\omega 
+\frac{1}{2}\langle\bar qq\rangle-
{1\over 32T^2}m_0^2 \langle\bar qq\rangle
-{g^2<({\bar q}\gamma_\mu {\lambda^a\over 2}q)^2>\over 2304T^3}
\;,
\end{equation}
where $m_0^2\,\langle\bar qq\rangle=\langle\bar qg\sigma_{\mu\nu}G^{\mu\nu}q\rangle$.

Using the following standard values for the condensates
\begin{eqnarray}
\label{parameter}
\langle\bar qq\rangle&=&-(0.225 ~\mbox{GeV})^3\;,\nonumber\\
\langle\alpha_s GG\rangle&=&0.038 ~\mbox{GeV}^4\;,\nonumber\\
m_0^2&=&0.8 ~\mbox{GeV}^2\;,
\end{eqnarray}
and with $s_0 =(1.5\pm 0.1)$ GeV, $T=0.4 \sim 0.6$ GeV we get 
from  (\ref{mass-0})
\begin{equation}
\label{result1}
\bar\Lambda_{+,{1\over 2}}=(1.15\pm 0.10) ~~\mbox{GeV},
\end{equation}
\begin{equation}
\label{ff3}
f_{+, 1/2}=(0.570\pm 0.08) ~~\mbox{GeV}^{3/2} .
\end{equation}

For comparison here we also write down the sum rules for the $(1^+, 2^+)$ doublet.
\begin{equation}
\label{mass-2}
{1\over 2}f_{+,3/2}^2e^{-{\bar\Lambda}_{+,{3\over 2}}/{T}}=
{1\over 2\pi^2}\int_0^{s_0}\omega^4e^{-\omega/{T}}d\omega
-\frac{1}{12}\:m_0^2\:\langle\bar qq\rangle
-{1\over 16}\langle{\alpha_s\over\pi}G^2\rangle T\;,
\end{equation}
where the following interpolating currents for the $(1^+, 2^+)$ doublet are used
\begin{eqnarray}
\label{curr5}
J^{\dag\alpha}_{1,+,{3\over 2}1}&=&\sqrt{\frac{3}{2}}
\:\bar h_v\gamma^5(-i)\left(
{\cal D}_t^{\alpha}-\frac{1}{3}\gamma_t^{\alpha}\not\!{\cal D}_t\right)q\;,\\
\label{curr6}
J^{\dag\alpha_1,\alpha_2}_{2,+,}&=&\:\bar h_v
\frac{(-i)}{2}\left(\gamma_t^{\alpha_1}{\cal D}_t^{\alpha_2}
+\gamma_t^{\alpha_2}{\cal D}_t^{\alpha_1}
-\frac{2}{3}\;g_t^{\alpha_1\alpha_2}\not\!{\cal D}_t\right)q\;.
\end{eqnarray}
These currents are also a factor $\sqrt{ 2}$ larger than those in \cite{zhu-prd}.

\subsection{Previous sum rules for pionic couplings}

Let us define the decay amplitudes of the doublet $(0^+, 1^+)$ in full QCD
\cite{col-95}
\begin{eqnarray}
\label{coup4}
 M(B'_0\to B(k) \pi (q))&=&I\sqrt{m_{B'_0} m_B}
 {m_{B'_0}^2-m_B^2 \over 2m_{B'_0}} g  \;,\\
\label{coup5}
 M(B_1'\to B^* (k)\pi (q))&=&I \{ \epsilon^*\cdot\eta g 
 \sqrt{m_{B'_1} m_{B^*}} {k\cdot q\over m_{B'_1}}
 + (k\cdot \eta )(q\cdot \epsilon^* ) F \} \;,
  \end{eqnarray}
where $I=\sqrt{2}$, $1$ for charged and neutral pion respectively. The structure
$F$ vanishes in the $m_Q\to \infty $ limit.

In HQET these amplitudes have the simple form 
\begin{eqnarray}
\label{coup6}
 M(B'_0\to B\pi)&=&I\;g' \;,\\
\label{coup7}
 M(B_1'\to B^*\pi)&=&I\;\epsilon^*\cdot\eta g' \;
  \end{eqnarray}
where 
\begin{equation}
g'=-g( \bar\Lambda_{+,{1\over 2}}-\bar\Lambda_{-,{1\over 2}}) \;.
\end{equation}

For deriving the sum rules for $g'$ we consider the correlator
\begin{equation}\label{cor-2}
 \int d^4x\;e^{-ik\cdot x}\langle\pi(q)|T\{ J_{0,+,\frac{1}{2}}(0),
 J^{\dagger}_{0,-,\frac{1}{2}}(x) \}|0\rangle=
I\; G_{B_0^{\prime} B} (\omega,\omega')\;,
\end{equation}
where $k^{\prime}=k+q$, $\omega=v\cdot k$, $\omega^{\prime}=v\cdot
k^{\prime}$, $q^2=0$ and $J^{\dag}_{0,-,{1\over 2}}=\bar h_v\gamma_5 q$.

Again after invoking the naive quark-hadron duality we have the double 
dispersion relation:
\begin{equation}
\label{pole}
{f_{-,{1\over 2}}f_{+,{1\over 2}}g'\over 4 (\bar\Lambda_{-,{1\over 2}}
-\omega')(\bar\Lambda_{+,{1\over 2}}-\omega)}+{c\over \bar\Lambda_{-,{1\over 2}}
-\omega'}+{c'\over \bar\Lambda_{+,{1\over 2}}-\omega} 
+\mbox{continuum}\; .
\end{equation}

Expressing (\ref{cor-2}) with the pion wave functions \cite{bely-95}
we obtain the LCQSR for $g'$ \cite{zhu-epjc}:
\begin{equation}\label{final}
 g' f_{-,{1\over 2} } f_{+, {1\over 2} } = 2 F_{\pi}
 e^{ { \Lambda_{-,{1\over 2} } +\Lambda_{+,{1\over 2} } \over 2T }}
 \{ -\varphi_\pi^{\prime} (u_0) T^2f_1({s_0\over T}) +
\mu_\pi \varphi_P (u_0)T f_0({s_0\over T}) + g'_1(u_0)
\}\;,
\end{equation}
where $F_\pi =93$MeV, $\mu_\pi =-{<{\bar q} q>\over F_\pi^2}=1.32$GeV at the 
scale $\mu =1$GeV, $\varphi_P (u)$ etc are the light cone pion
wave functions defined by \cite{bely-95}
\begin{equation}
<\pi (q) |{\bar d} (x) i\gamma_5 u(0) | 0> =
\sqrt{2} F_\pi \mu_\pi \int_0^1 du e^{iuqx}\varphi_P (u) \;.
\end{equation}
$\varphi_\pi^{\prime} (u_0)$, $g'_1(u_0)$ are the first derivatives of
$\varphi_\pi (u)$, $g_1(u)$ at $u=u_0$, 
$u_0={T_1\over T_1+T_2}$, $T={T_1T_2\over T_1+T_2}$,
$T_1$, $T_2$ are the Borel parameters. 
We choose $T_1 =T_2$. Therefore $u_0 ={1\over 2}$. At this point 
$\varphi_\pi^{\prime} (u_0)$ and $g'_1(u_0)$ vanish.
The factor 
$f_n(x)=1-e^{-x}\sum\limits_{k=0}^{n}{x^k\over k!}$ is used to subtract the 
parton-like continuum contribution with the continuum threshold $s_0$.

Correcting a numerical error in \cite{zhu-prd} we obtain from (\ref{final})
\begin{equation}\label{num}
g' =3.6\pm 0.7 \; .
\end{equation}

\section{New sum rules with $B\pi $ intermediate states}

\label{sec3}
\subsection{Mass sum rules}
We have the dispersion relation for (\ref{cor-1})
\begin{equation}\label{dip-1}
\Pi ( \omega )={1\over \pi} \int {\rho (s)\over s- \omega  -i\epsilon }ds\;,
\end{equation}
where $\rho (s)$ is the spectral density in the limit $m_Q \to \infty$.
At the quark level,
\begin{equation}\label{d-q-1}
\rho_q (s) ={N_c\over 2\pi} s^2\; ,
\end{equation}
where $N_c =3$ is the color number.

Due to the spontaneous chiral symmetry breaking, there exist $(N_f^2 -1)$ 
massless Goldstone bosons, where $N_f$ is the light quark flavor number.
The S-wave combination $B\pi$ has the same quantum numbers as the $B'_0$ 
meson. So the interpolating current (\ref{curr1}) "sees" 
both $B\pi$ and $B'_0$. In other words, the contribution due to $B\pi$
intermediate states should be included explicitly when we write the spectral 
density at the phenomenological side. Otherwise the $B'_0$ pole contribution 
will be overestimated leading to an abnormal large residue. 
\begin{equation}\label{d-p-1}
\rho (s) ={\pi \over 2}f^2_{+,{1\over 2}} \delta (s- {\bar \Lambda_{+,{1\over 2}}} )
+\rho_\pi (s) +\cdots \; ,
\end{equation}
where the first term is the $B'_0$ pole and the $\rho_\pi (s)$ is the $B\pi$
intermediate states contribution. The excited states and the continuum contribution
is denoted by the ellipse.

We start from the full QCD Lagrangian to derive $\rho_\pi (s)$.
It was shown in \cite{blok} that
\begin{eqnarray}\label{rho_pi} \nonumber
\rho_\pi (p^2) =&\int {d^4k\over (2\pi )^3}\int {d^4q\over (2\pi )^3}
\theta (k_0) \theta (q_0) \delta (k^2 -m_B^2) \delta (q^2) \\ \nonumber
&(2\pi )^3 \delta ( {\vec p} -{\vec k} -{\vec q}) 
\pi \delta (p_0 -k_0 -q_0) 
\sum |\langle 0|j_{B'_0} (0) |B\pi \rangle |^2 \\ \nonumber
&={1\over 16\pi} (1 -{m_B^2\over p^2} ) 
\sum |\langle 0|j_{B'_0} (0) |B\pi \rangle |^2 \\
&={1\over 16\pi}(1 -{m_B^2\over p^2} ) 
{f_B^2\over f_\pi^2} ({m_B^2\over m_b +m_q})^2 (N_f -{1\over N_f})\;,
\end{eqnarray}
where the sum is over all the possible Goldstone bosons, 
$f_\pi =132$MeV, $j_{B'_0} =\bar q b$ is the interpolating current of
$B'_0$ in full QCD, $f_B$ is defined as,
\begin{equation}
< 0|{\bar B} \gamma_\mu\gamma_5 q|B (k)>=if_B k_\mu \;.
\end{equation}
In the last step of deriving (\ref{rho_pi}) the soft pion 
theorem has been used to calculate the matrix element.
\begin{equation}\label{soft}
<0|\bar b q |B(k) \pi (q)>_{q\to 0} =
{1\over f_\pi} <0|\bar b i\gamma_5 q |B(k)>=
-{f_B\over f_\pi} {m_B^2\over m_b +m_q} \;.
\end{equation}
Moreover the chiral limit $m_q\to 0$ has been used for all the light
quarks and the $SU(N_f)$ flavor symmetry has been used to relate the 
amplitudes for different Goldstone bosons. 
The factor $(N_f -{1\over N_f})$ in (\ref{rho_pi}) is the result of
summing over these states. 

Letting $m_B =m_b +{\bar \Lambda}_{-,{1\over 2}}$, $p^2 =m_b^2 +2m_bs $
and taking the heavy quark limit, i.e., $m_b\to \infty$, we have
\begin{equation}\label{rho_pi-1}
\rho_\pi (s)={1\over 8\pi }({f_{-,{1\over 2}}\over f_\pi })^2 
(s-{\bar \Lambda}_{-,{1\over 2}} ) 
\theta (s-{\bar \Lambda}_{-,{1\over 2}} ) 
\theta ({\bar \Lambda}_{+,{1\over 2}} -s)(N_f -{1\over N_f}) \;,
\end{equation}
where we have used the relation $f_{-,{1\over 2}}=\sqrt{m_b}f_B$ in 
the leading order of $1/m_Q$.
Note $\rho_\pi (s)$ is a linear function of $s$ while the free 
parton level spectral density $\rho_q (s)$ in 
(\ref{d-q-1}) is quadratic in $s$. 
The latter is suppressed compared with the former in the 
lower energy region. So the spectral density is significantly disturbed 
by the presence of light Goldstone bosons. In (\ref{rho_pi-1}) 
we have introduced the factor $\theta ({\bar \Lambda}_{+,{1\over 2}} -s)$.
The reason is that the soft pion theorem does not hold any more 
beyond the region $|\vec q|_\pi < {\bar \Lambda}_{+,{1\over 2}}-
{\bar \Lambda}_{-,{1\over 2}}\sim 350$MeV. 
Moreover it was conjectured in \cite{blok} that $<0|{\bar b}q|B\pi >$
drops when the total energy of $B\pi$ becomes larger than 
the mass of $B'_0$ so that the quark-hadron duality is restored
after integrating the energy over a larger interval from 
zero to the continuum threshold.

Note $m_K =498$MeV and $m_\eta =547$MeV due to nonzero current quark mass.
So in realistic case only $B\pi$ intermediate states 
contribute to $\rho_\pi (s)$ in (\ref{rho_pi-1}) corresponding to $N_f =2$.  
Now we arrive at the new sum rules after making Borel transformation:
\begin{eqnarray}
\label{mass-1}\nonumber 
{1\over 2}f_{+,1/2}^2e^{-{\bar\Lambda_{+,{1\over 2}}/T}} +
{3\over 16\pi^2 }({f_{-,{1\over 2}}\over f_\pi })^2 
\int_{{\bar \Lambda}_{-,{1\over 2}}}^{{\bar \Lambda}_{+,{1\over 2} }}
(s-{\bar \Lambda}_{-,{1\over 2}} )e^{-s/T}ds &\\ 
=\frac{3}{2\pi^2}\int_0^{s_0}s^2 e^{-s/T}ds+\frac{1}{2}
\langle\bar qq\rangle-{1\over 32T^2}m_0^2 \langle\bar qq\rangle
-{g^2<({\bar q}\gamma_\mu {\lambda^a\over 2}q)^2>\over 2304T^3}
\;, &
\end{eqnarray}
where $s_0$ is the continuum threshold. Starting from $s_0$ we have modeled
the phenomenological spectral density with the free parton-like one.

\subsection{New sum rules for $g'$}

Similarly we can write the double dispersion relation 
in the leading order of HQET for (\ref{cor-2}) as
\begin{equation}\label{dip-2}
\Pi (\omega , \omega' )={1\over \pi^2} \int {\rho (s, s')\over
(s-\omega ) (s'-\omega') } ds ds' +\cdots\;,
\end{equation}
where the ellipse denotes the subtraction terms.

The pole term is
\begin{equation}
\rho_p (s,s' )={\pi^2\over 4} g' f_{+,{1\over 2}}f_{-,{1\over 2}}
\delta (s -{\bar \Lambda}_{-,{1\over 2}} )
\delta (s' -{\bar \Lambda}_{+,{1\over 2}} ) \;.
\end{equation}
and the contribution of the $B\pi$ intermediate states in full QCD is
\begin{eqnarray}\label{rho_pi-3} \nonumber
&\rho_\pi (s, s') =\int {d^4k\over (2\pi )^3}\int {d^4 l\over (2\pi )^3}
\theta (k_0) \theta (l_0) \delta (k^2 -m_B^2) \delta (l^2) \\ 
&(2\pi )^3 \delta ( {\vec p} -{\vec k} -{\vec l}) 
\pi \delta (p_0 -k_0 -l_0) 
\sum |\langle \pi (q) |j_{B} (0) |B(k)\pi (l) \rangle 
\langle B(k)\pi (l) |j_{B'_0} (0) |0 \rangle|
\;.
\end{eqnarray}
Using the soft pion limit and $SU_f(2)$ symmetry we find
\begin{equation}\label{wsx}
\rho_\pi (s, s') ={3\over 32}{f_Bf_{B'_0}g'\over f_\pi^2} 
{m_B^2\over m_b +m_q} 
(1 -{m_B^2\over s'} ) \delta (s-m_{B'_0}^2 ) \;.
\end{equation}
Taking the heavy quark limit (\ref{wsx}) is reduced to
\begin{equation}
\rho_\pi (s,s' )={3\over 32 f_\pi^2}g' f_{+,{1\over 2}}f_{-,{1\over 2}}
(s' -{\bar \Lambda}_{-,{1\over 2}} ) 
\theta (s' -{\bar \Lambda}_{-,{1\over 2}} )
\theta ({\bar \Lambda}_{+,{1\over 2}} -s' )
\delta (s -{\bar \Lambda}_{+,{1\over 2}} ) \;.
\end{equation}

Finally we have a new sum rule for $g'$:
\begin{eqnarray}\label{final-1}\nonumber
 g' f_{-,{1\over 2} } f_{+, {1\over 2} } =&  2 F_{\pi}
 e^{ { \Lambda_{+,{1\over 2} } \over 2T }}
\{  e^{ -{ \Lambda_{-,{1\over 2} } \over 2T }}
+ {3\over 8\pi^2f_\pi^2}
\int_{ {\bar \Lambda}_{-,{1\over 2}}}^{ {\bar \Lambda}_{+,{1\over 2}}}
(s'-{\bar \Lambda}_{-,{1\over 2}} )e^{-{s' \over 2T}} ds' \}^{-1}\\
& \{ -\varphi_\pi^{\prime} (u_0) T^2f_1({s_0\over T}) +
\mu_\pi \varphi_P (u_0)T f_0({s_0\over T}) + g'_1(u_0)
\}  \;. 
\end{eqnarray}

\subsection{Numerical analysis}
\label{num-33}
As input we need $\bar\Lambda_{-,1/2}=0.5$ GeV and 
$f_{-,1/2}\simeq 0.35$ GeV$^{3/2}$ at the order $\alpha_s=0$ \cite{neubert}.
The numerical results for $\bar\Lambda_{+,1/2}, f_{+,1/2}$ and 
$\bar\Lambda_{+,3/2}, f_{+,3/2}$ in \cite{zhu-prd} were obtained 
by first applying the operator ${d\ln \over d(1/T)}$ to (\ref{mass-0})
and (\ref{mass-2}) to extract $\bar\Lambda_{+,1/2}$ and $\bar\Lambda_{+,3/2}$,
which were then used to obtain $f_{+,1/2}$ and $f_{+,3/2}$ respectively.
Here we use a different procedure which appears to be better. This 
involves with simultaneously varying the parameters ${\bar \Lambda}_{+,{1\over 2}}, 
f_{+,{1\over 2}}, s_0$ etc to find the best fitting of the left 
hand side (L.H.S.) and right hand side (R.H.S.) of the sum rules.
We work at the region $T>0.4$GeV for Eq. (\ref{mass-1}), 
where the power correction is under control. 
We allow the continuum threshold to vary from $1.06$ GeV to $1.46 $GeV.
Numerically we have
\begin{equation}\label{lambda-2}
{\bar \Lambda}_{+,{1\over 2}}=(0.85\pm 0.15)\mbox{GeV}\;,
\end{equation}
\begin{equation}\label{fff}
f_{+,{1\over 2}}=(0.36\pm 0.10)\mbox{GeV}^{3\over 2}\;.
\end{equation}
With these parameters the left hand side and right hand side agree 
within five percent in the region $0.5 < T < 0.8$GeV as can be 
seen from FIG. 1. 
Typically at $T=0.4$ GeV the sum of the $B'_0$ pole and $B\pi$ 
intermediate states constitutes about $60\%$ of the whole sum rule. 
The continuum starting from $s_0$ is about $40\%$.

It is important to notice that the $B\pi$ intermediate states contribute
about $15\%$ to the left hand side of (\ref{mass-1}). 
If we use this fitting method in the numerical 
analysis of the old sum rules (\ref{mass-0})
we reproduce the results in \cite{nucl-shuryak} 
\begin{equation}\label{lambda-3}
{\bar \Lambda}_{+,{1\over 2}}=(1.2\pm 0.2)\mbox{GeV}\;,
\end{equation}
\begin{equation}\label{fff-3}
f_{+,{1\over 2}}=(0.75\pm 0.15)\mbox{GeV}^{3\over 2}\;,
\end{equation}
\begin{equation}\label{sss-3}
s_0=(1.8\pm 0.2)\mbox{GeV}\;.
\end{equation}
Note in \cite{nucl-shuryak} no error is given for $f_{+,{1\over 2}}$.
The value and error in (\ref{fff-3}) is the result of our reanalysis.
We see that both $f_{+,{1\over 2}}$ and ${\bar \Lambda}_{+,{1\over 2}}$ 
are significantly reduced after taking into account $B\pi$ intermediate states.

We can also apply the fitting method to the analysis of the sum rules for the 
$(1^+, 2^+)$ doublet.
The fitted curve of the L.H.S.
and the parton level curve of the R.H.S. of Eq. (\ref{mass-2})
are shown in FIG. 2 in the region $T>0.4$GeV with the following most suitable 
parameters ${\bar \Lambda}_{+,{3\over 2}}, f_{+,{3\over 2}}, s_0$. 
\begin{equation}\label{lambda-4}
{\bar \Lambda}_{+,{3\over 2}}=(0.95\pm 0.10)\mbox{GeV}\;,
\end{equation}
\begin{equation}\label{ffff}
f_{+,{3\over 2}}=(0.263\pm 0.06)\mbox{GeV}^{3\over 2}\;,
\end{equation}
\begin{equation}\label{ssss}
s_0=(1.3\pm 0.2)\mbox{GeV}\;.
\end{equation}
With these parameters the fitting is excellent, typically with an
accuracy within one percent in the large interval 
$0.5 < T < 1.4$GeV as can be seen from FIG. 2. 

The central value of ${\bar \Lambda}_{+,{1\over 2}}$ 
is about $100$ MeV lower than that of ${\bar \Lambda}_{+,{3\over 2}}$, 
in good agreement with the experimental data \cite{L3}.

Now we are ready to extract $g'$. 
Using the same pion wave functions as in \cite{zhu-epjc} we have
\begin{equation}
\label{res}
g'f_{-,{1\over 2} } f_{+, {1\over 2} }
   =(0.36\pm 0.05)~~~\mbox{GeV}^{3}\;,\\
\end{equation}
where the error refers to the variations with $T$ and $s_0$. 
And the central value corresponds to $T=0.9$GeV and $s_0 =1.26$GeV. 
The variation of the left hand side of (\ref{final-1}) 
with $T$ and $s_0$ is presented in FIG. 2.
Finally we get
\begin{equation}\label{num-2}
g' =2.8\pm 0.5 \; .
\end{equation}

The decay width formulas in the leading order of $1/m_Q$ are
\begin{equation}\label{dd-1}
\Gamma(B'_0\to B\pi)={3\over 8\pi} g^{'2}|\vec q|_\pi\;,
\end{equation}
\begin{equation}\label{dd-2}
\Gamma(B'_1\to B^*\pi)={3\over 8\pi} g^{'2}|\vec q|_\pi \;.
\end{equation}
In order to include the large $1/m_Q$ correction in the kinematical factors,
we use the decay width formulas with finite $m_Q$ instead of (\ref{dd-1}),
(\ref{dd-2}).
\begin{equation}
\Gamma(B'_0\to B\pi)={3\over 32\pi} g^{2}{m_B (m_{B'_0}^2-m_B^2)^2\over 
m_{B'_0}^3 }|\vec q|_\pi\;,
\end{equation}
\begin{equation}
\Gamma(B'_1\to B^*\pi)={1\over 32\pi} g^{2}
{m_{B^*} (m_{B'_1}^2-m_{B^*}^2)^2\over m_{B'_1}^3 }
[2+{(m_{B'_1}^2+m_{B^*}^2)^2
\over 4m_{B'_1}^2m_{B*}^2 }]|\vec q|_\pi\;.
\end{equation}

Numerically we have
\begin{equation}
\Gamma(B'_0\to B\pi) \approx 250\mbox{MeV}\;,
\end{equation}
\begin{equation}
\Gamma(B'_1\to B^*\pi)\approx 250\mbox{MeV}\;,
\end{equation}
with $m_{B'_1}=m_{B^*}+350 \mbox{MeV}$, $m_{B'_0}=m_B+350 \mbox{MeV}$.

\section{Summary}
\label{sec4}

In summary, we have reanalyzed the QCD sum rules for both the $(0^+, 1^+)$
mass and its pionic decay amplitude. The contribution of 
the $B\pi$ intermediate states in the soft pion region 
are investigated in detail. The spectral density is disturbed significantly 
by the presence of Goldstone bosons. After subtracting the contribution of these 
intermediate states, we have obtained 
new results for the binding energy ${\bar \Lambda_{+,{1\over 2}}}$
and decay widths for the $(0^+, 1^+)$ doublet in the leading order of HQET. 
Numerically ${\bar \Lambda_{+,{1\over 2}}}=(0.85\pm 0.15)\mbox{GeV}$,
which is about $100$ MeV smaller than 
${\bar \Lambda_{+,{3\over 2}}}=(0.95\pm 0.10)\mbox{GeV}$ 
extracted with the same fitting method, in good
agreement with the most recent experimental data \cite{L3}. 
The $(0^+, 1^+)$ decay width is around 250 MeV, which
remains to be larger than the experimental result 
$(76\pm 28(\mbox{stat})\pm 15(\mbox{syst}))$ MeV in \cite{L3}. 
The origin of this discrepancy is not clear at present.
We want to point out that the same contamination from the Goldstone bosons
exists for the sum rules for the $(1^+, 2^+)$ doublet \cite{dyb-1,zhu-prd}. 
But in this case the $B\pi$ intermediate states disturb the spectral 
density only slightly for they are in the D-wave state. 

The errors for our numerical results given in Sec. \ref{num-33} include
only those from the variation of the Borel parameter $T$ and the 
continuum thresold $s_0$ in the window. They don't include those 
from the uncertainty of the condensates and intrinsic errors 
of the QCD sum rule approach. 
In our analysis we have neglected the contribution of $B\pi$ intermediate
states to the spectral density for $s> {\bar \Lambda}_{+,{1\over 2}}$ 
since the soft pion theorem does not hold any more and there is not a
reliable way to estimate the matrix element $<0|\bar b q |B(k) \pi (q)>$.
This is another source of uncertainty.
After we submitted the original version of this paper, we
learned that CLEO collaboration have measured the $D'_0$ mass and width 
to be $2461$ MeV and $200\sim 400$ MeV respectively \cite{cleo}.

\vspace{0.8cm} {\it Acknowledgments:\/} S.Z. was supported by
the Natural Science Foundation and Postdoctoral Science Foundation of China.
Y.B. was supported by the Natural Science Foundation of China.

\vspace{1.0cm}
{\bf Figure Captions}
\vspace{2ex}
\begin{center}
\begin{minipage}{130mm}
{\sf FIG. 1.} \small{The varaition of the right and left hand side 
of Eq. (\ref{mass-1}) with Borel parameter $T$ 
is plotted as solid and dotted curves respectively with the fitting 
parameters in (\ref{lambda-2})-(\ref{fff}).
}
\end{minipage}
\end{center}
\begin{center}
\begin{minipage}{130mm}
{\sf FIG. 2.} \small{The varaition of the right and left hand side 
of Eq. (\ref{mass-2}) with $T$ using the central 
values of the fitting parameters in (\ref{lambda-4})-(\ref{ssss}).
}
\end{minipage}
\end{center}
\begin{center}
\begin{minipage}{130mm}
{\sf FIG. 3.} \small{The sum rules for $g'f_{-,{1\over 2}}f_{+,{1\over 2}}$  
with $s_0=1.36, 1.26, 1.16$ GeV respectively. }
\end{minipage}
\end{center}

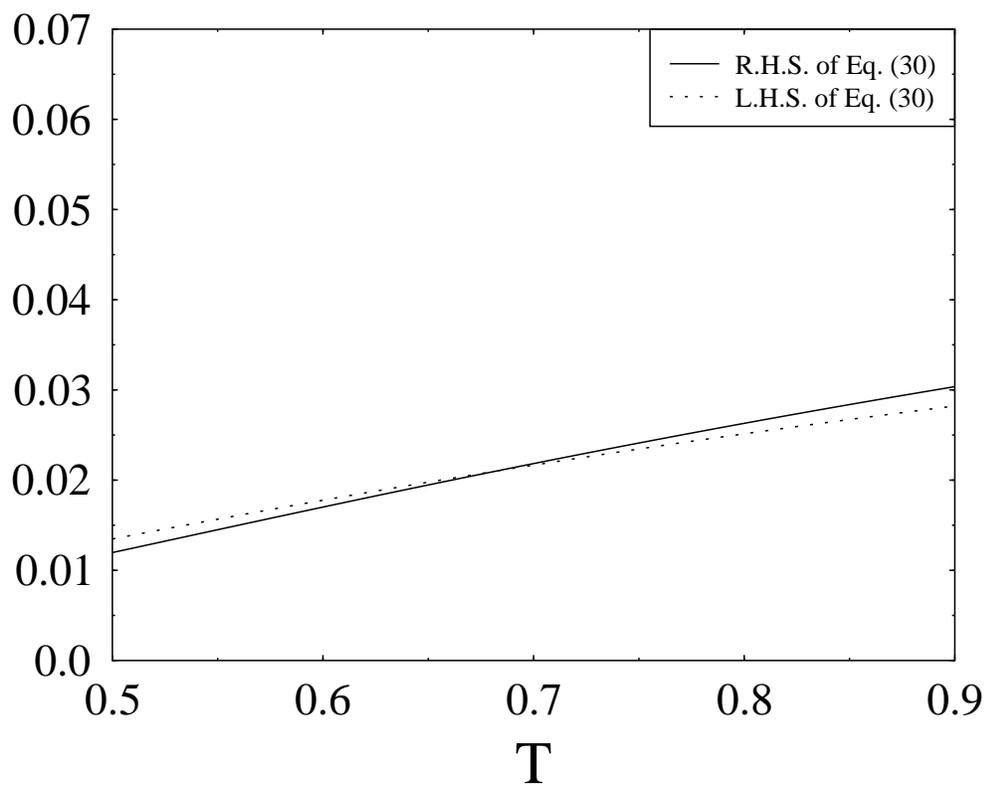
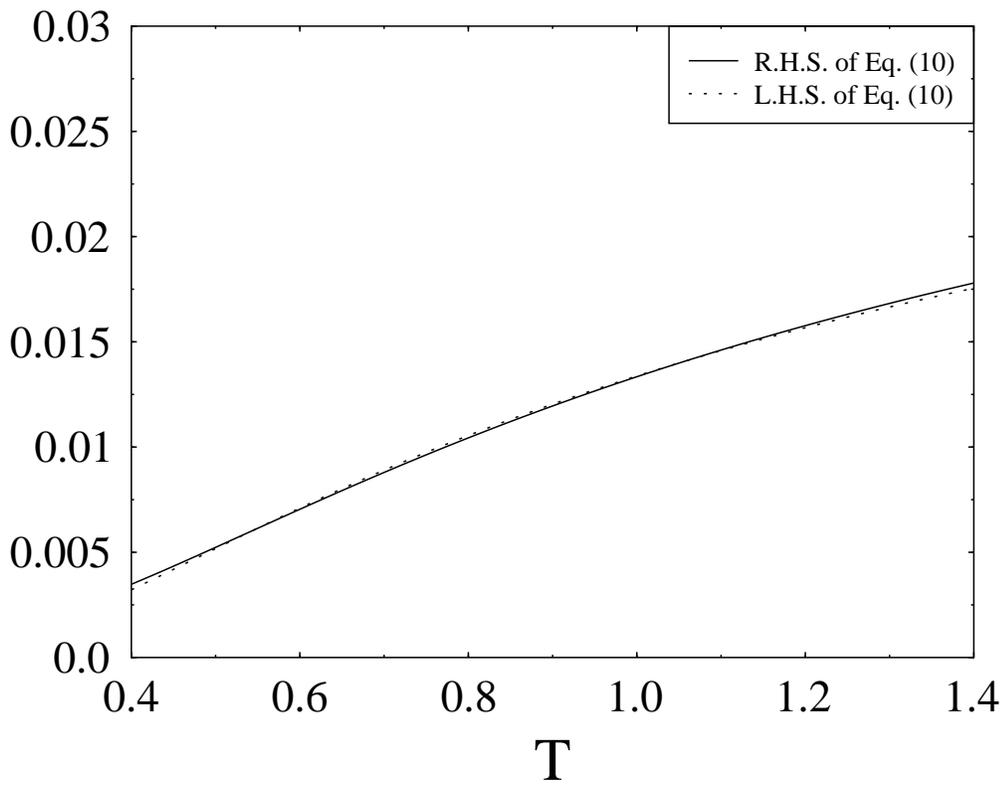
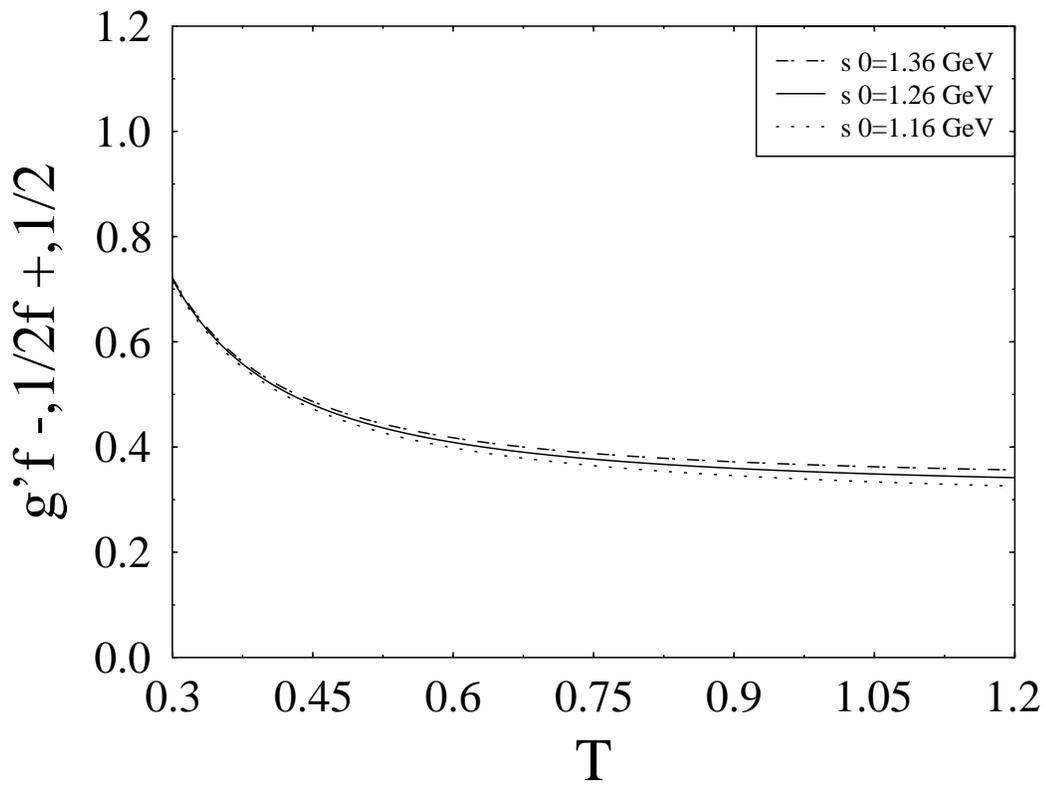
\end{document}